\documentclass[aps,preprint,showpacs]{revtex4}
\usepackage{amsfonts,amsmath,amssymb,bm,eucal}
\usepackage[english]{babel}
\usepackage[latin1]{inputenc}
\usepackage[T1]{fontenc}
\usepackage{ae,aecompl}
\usepackage{graphicx,graphics,epsfig,epstopdf,subfigure,color,psfrag}
\usepackage{hyperref}
\newcommand{\spur}[1]{\not\! #1 \,}
\newcommand{\be}{\begin{equation}}
\newcommand{\ee}{\end{equation}}
\newcommand{\bea}{\begin{eqnarray}}
\newcommand{\eea}{\end{eqnarray}}

\newcommand{\dd}{\displaystyle}

\begin{document}

\preprint{BARI-TH/652-13}
\title{ Bounds on  the compactification scale of  two universal extra dimensions from exclusive $b \to s \gamma$ decays\\}
\author{Pietro Biancofiore }
\affiliation{
Dipartimento di Fisica,  Universit$\grave{a}$  di Bari, Italy\\
Istituto Nazionale di Fisica Nucleare, Sezione di Bari, Italy\\}

\begin{abstract}

The exclusive radiative  $B \to K^* \gamma$, $B \to K^*_2 \gamma$, $B_s \to \phi \gamma$ and $\Lambda_b \to \Lambda \gamma$  decays are studied in a new physics  scenario with two universal extra dimensions compactified on a chiral square. The computed branching fractions depend on the size $R$ of the extra dimensions,  and a comparison with the available measurements allows us to put bounds on such a fundamental parameter. From the mode $B^0 \to K^{*0} \gamma$  we obtain  the most stringent bound: $\displaystyle{1 \over R}>710$ GeV.
\end{abstract}

 \maketitle

Among the various new physics (NP) scenarios  proposed to extend the standard model (SM), those with extra dimensions (ED) are particularly interesting, as they might be  able to solve some of the problems  affecting the SM without invoking  the existence of new interactions,  changing the geometry of the space-time. Indeed,  although the attempt of T. Kaluza and O. Klein, as well as of G. Nordstr\"om,  to unify  the electromagnetism with gravity  introducing one  extra dimension was  unsuccessful \cite{KK},  the idea of the possible  existence of additional space-like dimensions resulted to be fruitful, and has been pursued for sectors of fundamental interactions with the formulation of different settings involving ED.  Such models can provide a unified framework for gravity and other interactions,   hints on the hierarchy problem, connections with the string theory,  interesting dark matter phenomenology and, remarkably, they can  be at the origin of observable phenomena  at the colliders and at the 
flavour factories \cite{rev}. 
A common feature  is that the compactification of the extra dimensions implies  the existence of Kaluza-Klein (KK) partners of  SM fields in the four-dimensional description of the higher dimensional theory,  together with  KK modes without SM correspondents.  Models in which all the SM fields  propagate in the extra dimensions are usually denoted as universal extra dimension (UED) scenarios.

A simple scenario is  the Appelquist, Cheng and Dobrescu (ACD) model \cite{Appelquist:2000nn},  a minimal extension of SM in $4 + 1$ dimensions, with the extra dimension compactified to the orbifold $S^1/Z_2$ and the fifth coordinate y running from 0 to $2\pi R$, $y=0$ and $y=\pi R$ being fixed points of the orbifold. In this model  the SM particles correspond to  the zero modes of fields propagating in the compactified extra dimension, and the choice of  the extra dimension topology is dictated by the need of  having  chiral fermion zero modes.
 In addition to the zero modes,    towers of KK excitations are predicted to exist, corresponding to the heavier modes of the fields in the extra dimension; such fields  are imposed to be even under a parity transformation in the fifth coordinate $P_5: y \to
-y$.  On the other hand,  fields  which are odd under $P_5$ propagate in the extra dimension without zero modes, and correspond to particles  without  SM partners.
The masses of  KK particles  depend on  the radius $R$ of the  extra dimension orbifold, the  only new parameter with respect to SM.
  For example,  the  masses of  KK bosonic modes are given by $ m_n^2=m_0^2+\displaystyle{n^2 \over R^2} \,\,\,\,\,\,\, n=1,2,\dots $, with $m_0$  the mass of the zero mode, so that   for small values of $R$, i.e. at large compactification scales,  the KK particles decouple from the low energy sector.
 Another  property of the ACD model is  the conservation of the KK parity $(-1)^j$,  $j$ being  the KK number. This prevents tree level contributions of Kaluza-Klein states to low energy ($\mu \ll 1/R$) processes, forbidding the production of a single KK particle  off the interaction of standard particles.
As a consequence, precise electroweak measurements can be used  to derive a lower bound to the compactification scale: ${1 / R} \ge 250-300$ GeV \cite{Appelquist:2002wb}.
 Moreover, this  leads to  the possibility that the lightest  KK particles, for instance the $n=1$ Kaluza-Klein  excitations of the photon and  of the neutrinos,  are among  the  dark matter components  \cite{Cheng:2002iz,Hooper:2007qk}.
KK modes could be  produced at colliders,  or indirectly  detected through investigations of    loop-induced processes getting contributions from  KK modes. This possibility has been explored considering  Flavour Changing Neutral Current (FCNC) transitions in the ACD scenario \cite{UED1,UED1b,Colangelo:2007jy},   with a  stringent constraint  on $\displaystyle{1/ R}$  obtained from the exclusive radiative $B \to K^* \gamma$ decay \cite{Colangelo:2006} and from the inclusive $B \to X_s \gamma$ mode \cite{Haisch}.

Models with universal extra dimensions may have a larger number of additional  dimensions,  $d\geq 2$, with  further theoretical motivations than for $d=1$. For instance,  a model with  two  compactified UEDs  allows to cancel the global SU(2) anomaly with three generations \cite{Dobrescu:2004zi}; the compactification of two universal extra dimensions on the so-called  ``chiral square''  induces the suppression of the proton decay rate even with maximal violation of the baryon number at the TeV scale, as a result of the preservation of a discrete symmetry (which is the generator of a subgroup of the six dimensional Lorentz group)  \cite{Appelquist:2001mj}.
This is the model proposed in \cite{Dobrescu:2004zi} and  considered here:  we denote it as  6UED.
The two extra dimensions are flat and compactified on a  square of side $L$: $0 \le x^4,x^5 \le L$, where $x^4$ and $x^5$ are  the fifth and sixth extra spatial coordinates. The compactification is performed identifying two pairs of adjacent sides of the square: $(y,0)=(0,y)$ and $(y,L)=(L,y)$,  for all $y \in [0,L]$,  which amounts to folding the square along a diagonal. The fields are decomposed in Fourier modes,  in terms of effective four dimensional fields labeled by two indices $(l,k)$. Hence,  the KK modes are identified by two KK numbers which determine their mass in four dimensions:  zero modes correspond to SM fields.

The values of the fields in the points identified through the folding are related by a symmetry transformation. For instance, for a scalar field,  the field values may differ by a phase. The choice of the folding boundary conditions (and of the constraints on such phases) is mostly important in the case of fermions, since a suitable choice allows to obtain chiral zero modes, while higher KK modes have masses determined (as for scalars) by the relation: $M_{l,k}=\displaystyle{\sqrt{l^2+k^2} \over R}$, where $R=\displaystyle{L / \pi}$ is now the compactification radius, with $l$ and $k$ integer numbers.
The theory has an additional symmetry, the  invariance under reflection with respect to the center of the square,  which distinguishes between the various KK excitations of a given particle. A KK mode identified by the pair $(l,k)$ of indices changes sign under reflection if $l+k$ is odd, while it remains invariant if $l+k$ is even. As a consequence,  stability of the lightest KK modes is guaranteed, so that such  are good candidates for dark matter constituents.

The model describes  SM particles, their KK excitations, and new particles without a SM correspondent  from fields whose Fourier decomposition does not contain a zero mode. As an example, two scalar fields originate from the mixing of the fourth and the fifth component of the vector gauge bosons. Then, at each KK level, a linear combination due to a further mixing of these two fields with the Higgs field is promoted to a pseudo-Goldstone mode reabsorbed to give mass to the spin-1 KK mode; another combination named the ``spinless adjoint'' \cite{Dobrescu:2001ae} remains as a physical real scalar field. The consequence is an interesting phenomenology which manifests itself in the production cross sections and decay modes of other KK states,  which differ from the $d=1$ case: as an example, the (1,1) states with mass $\sqrt{2}/R$,  can be produced in the $s-$channel \cite{Allanach:2006fy, Dobrescu:2004zi, Burdman:2006gy}.
The phenomenology of KK excitations has been investigated in view of collider searches  \cite{Dobrescu:2007xf,Ghosh:2008ji} as well as for the  cosmological and dark matter implications \cite{Dobrescu:2007ec,Bertone:2010ww}.

The new particles may contribute as virtual states to  FCNC processes,  and actually from
 the inclusive $B \to X_s \gamma$ decay   the bound $\displaystyle{1 \over R} \ge 650$ GeV (at 95$\%$ C.L.)  has been obtained \cite{Freitas:2008vh}.
The  analysis can be extended to the  exclusive $b \to s \gamma$ induced modes, prompted by the example of a single UED in which such processes turned out to be the effective in constraining the compactification scale. Here we focus on the transitions $B \to K^* \gamma$, $B \to K_2^*(1430) \gamma$, $B_s \to \phi \gamma$ and $\Lambda_b \to \Lambda \gamma$.  For some of these processes, precise  experimental data are  available and can be exploited to put bounds on $\displaystyle{1/R}$. The modes that have not been observed, yet, 
can be explored at the CERN LHC and at the new planned flavour factories.

In  SM the effective weak Hamiltonian describing the
 $b \to s \gamma$ and $b \to s \, {\rm gluon}$ transition can be written as \cite{Buchalla:1995vs}:
\be
H_{eff}=-{G_F \over \sqrt{2}} V_{tb}V_{ts}^* \left( \sum_{i=1}^6 C_i \, O_i+C_{7 \gamma} \, O_{7 \gamma}+C_{8G} \,  O_{8G} \right) \,. \label{heff}
\ee
$G_F$ is the Fermi constant and $V_{ij}$ are elements of the Cabibbo-Kobayashi-Maskawa (CKM) mixing matrix;  terms proportional to $V_{ub} V_{us}^*$ are neglected in (\ref{heff}) since the ratio
$\displaystyle \left |{V_{ub}V_{us}^* \over V_{tb}V_{ts}^*}\right|$ is ${O}(10^{-2})$.
$C_i$ are Wilson coefficients, and $O_i$ are local operators written in terms of quark and gluon fields:
\begin{eqnarray}
O_1&=&({\bar s}_{L \alpha} \gamma^\mu b_{L \alpha})
      ({\bar c}_{L \beta} \gamma_\mu c_{L \beta}) \nonumber \\
O_2&=&({\bar s}_{L \alpha} \gamma^\mu b_{L \beta})
      ({\bar c}_{L \beta} \gamma_\mu c_{L \alpha}) \nonumber \\
O_3&=&({\bar s}_{L \alpha} \gamma^\mu b_{L \alpha})
      [({\bar u}_{L \beta} \gamma_\mu u_{L \beta})+...+
      ({\bar b}_{L \beta} \gamma_\mu b_{L \beta})] \nonumber \\
O_4&=&({\bar s}_{L \alpha} \gamma^\mu b_{L \beta})
      [({\bar u}_{L \beta} \gamma_\mu u_{L \alpha})+...+
      ({\bar b}_{L \beta} \gamma_\mu b_{L \alpha})] \nonumber \\
O_5&=&({\bar s}_{L \alpha} \gamma^\mu b_{L \alpha})
      [({\bar u}_{R \beta} \gamma_\mu u_{R \beta})+...+
      ({\bar b}_{R \beta} \gamma_\mu b_{R \beta})]  \\
O_6&=&({\bar s}_{L \alpha} \gamma^\mu b_{L \beta})
      [({\bar u}_{R \beta} \gamma_\mu u_{R \alpha})+...+
      ({\bar b}_{R \beta} \gamma_\mu b_{R \alpha})] \nonumber \\
O_{7 \gamma}&=&{e \over 16 \pi^2} \left[ m_b ({\bar s}_{L \alpha}
\sigma^{\mu \nu}
     b_{R \alpha}) +m_s ({\bar s}_{R \alpha}
\sigma^{\mu \nu}
     b_{L \alpha}) \right] F_{\mu \nu} \nonumber \\
O_{8G}&=&{g_s \over 16 \pi^2} m_b \Big[{\bar s}_{L \alpha}
\sigma^{\mu \nu}
      \Big({\lambda^a \over 2}\Big)_{\alpha \beta} b_{R \beta}\Big] \;
      G^a_{\mu \nu} \,\,; \nonumber
\label{eff}
\end{eqnarray}
\noindent
$\alpha$, $\beta$ are color indices, $\displaystyle
b_{R,L}={1 \pm \gamma_5 \over 2}b$, and $\displaystyle \sigma^{\mu
\nu}={i \over 2}[\gamma^\mu,\gamma^\nu]$; $e$ and $g_s$ are the
electromagnetic and the strong coupling constant, respectively, $m_b$ and $m_s$ the beauty and the strange quark masses, 
$F_{\mu \nu}$  in $O_{7 \gamma}$ and $G^a_{\mu \nu}$ in  $O_{8 \gamma}$ denote the
electromagnetic and the gluonic field strength tensors, and $\lambda^a$ are the Gell-Mann matrices.

  The Wilson coefficients  in (\ref{heff}) have been computed at NNLO in SM \cite{nnlo}.
The most relevant contribution to $b \to s \gamma$ comes from the operator $ O_{7 \gamma}$, a magnetic penguin specific of such a transition. The term proportional to $m_s$ contributes much less than the one proportional to $m_b$, the reason for which the emission of left-handed photons dominates over that of right handed ones in  SM. Since  the coefficient  $C_{7 \gamma}$ depends on the  regularization scheme,  it is convenient to consider, at leading order, a combination that is  regularization scheme independent \cite{Buras:1993xp}:
\begin{equation}
 C_{7\gamma}^{(0)eff}(\mu_b)=\eta^{16 \over 23}
C_{7\gamma}^{(0)}(\mu_W)+{8 \over 3} \left( \eta^{14 \over 23} -\eta^{16
\over 23} \right)C_{8G}^{(0)}(\mu_W)+C_2^{(0)}(\mu_W) \sum_{i=1}^8
h_i \eta^{a_i} \,\,\, , \ee
where $\displaystyle \eta={\alpha_s(\mu_W) \over
\alpha_s(\mu_b)}$ and $C_2^{(0)}(\mu_W)=1$
(the superscript $(0)$ stays for leading log approximation), with
\bea 
a_1={14 \over 23} \,\,\, , \hskip 0.5cm a_2={16 \over 23} \,\,\, ,
\hskip 0.4cm && a_3={6 \over 23} \,\,\, ,\hskip 0.5cm a_4=-{12 \over 23} \,\,\, ,
\nonumber \\ a_5= 0.4086 \,\,\, ,\hskip 0.5cm a_6=-0.4230 \,\,\, ,\hskip 0.5cm &&
a_7=-0.8994 \,\,\, ,\hskip 0.5cm
a_8=0.1456 \,\,\, ,\nonumber \\
h_1=2.2996 \,\,\, ,\hskip 0.5cm h_2=-1.0880  \,\,\, ,\hskip 0.5cm &&
 h_3=-{3 \over 7}  \,\,\, ,\hskip 0.5cm  h_4=-{1 \over 14} \,\,\, ,\label{numbers}
 \\
 h_5=-0.6494 \,\,\, , \hskip 0.5cm h_6=-0.0380 \,\,\, ,\hskip 0.5cm && h_7= -0.0185 \,\,\, , \hskip 0.5cm
 h_8=-0.0057 \,\,\, . \nonumber \eea

The ACD and the 6UED models belong to the class of minimal flavour violation models, therefore the only modification  with respect to  the SM consists in a different value of the Wilson coefficients in the effective weak Hamiltonian (\ref{heff}), without new operators. The explicit expression of $C_{7\gamma}^{eff}$ in the case of two extra dimensions can be found  in Ref. \cite{Freitas:2008vh}. It should only be mentioned that the sums over the KK modes entering in the expression of the Wilson coefficients in the extra dimensional framework diverge logarithmically, and should be
 cut in correspondence of some values of $N_{KK}=l+k$, viewing  this theory  as an effective one valid up to a some higher scale. Following \cite{Freitas:2008vh}, the  condition  $N_{KK} \simeq 10$ can be chosen.

 To consider the contribution of the effective weak vertex $O_{7 \gamma}$ to the transitions of interest here  we need  the  hadronic matrix elements
 \bea
<V(p^\prime,\eta)|{\bar s} \sigma_{\mu \nu} q^\nu
  b |B_{(s)}(p)>&=& i \epsilon_{\mu \nu \alpha
\beta}\, \eta^{* \nu} p^\alpha (p^{\prime})^\beta
\; 2 \; T_1^{B_{(s)} \to V}(q^2)   \label{t1BK*}
\\
<V(p^\prime,\eta)|{\bar s} \sigma_{\mu \nu} q^\nu
 \gamma_5  b |B_{(s)}(p)>&=&
  \Big[ \eta^*_\mu (M_{B_{(s)}}^2 - M_V^2)  -
(\eta^* \cdot q) (p+p^\prime)_\mu \Big] \; T_2^{B_{(s)} \to V}(q^2) \nonumber \\
&+& (\eta^* \cdot q) \left [ q_\mu - {q^2 \over M_{B_{(s)}}^2 -
M_V^2} (p + p^\prime)_\mu \right ] \; T_3^{B_{(s)} \to V}(q^2)  \,\,\, .   \label{t23BK*}
\end{eqnarray}
$V$ stays for $K^*$ and $K_2^*(1430)$ in the case of $B$ decays,  and for $\phi(1020)$ in the case of $B_s$ decays; $q=p-p^\prime$ is the photon momentum and
 $\eta$  the  $V$ meson polarization vector. In the case of $K^*_2(1430)$, which is a spin 2 particle, the polarization vector is described by a two indices symmetric and traceless tensor, therefore in (\ref{t1BK*}-\ref{t23BK*})    $\eta^\alpha=\eta^{\alpha \beta}\displaystyle{p_\beta \over M_B}$. Relations exist among the three form factors $T_i^{B \to V}$, $i=1,2,3$, in particular the condition: $T_1^{B_{(s)} \to V}(0) =T_2^{B_{(s)} \to V}(0) $.
 Due to this relation, the rate of the processes $B_{(s)}(p) \to V(p^\prime,\, \eta) \, \gamma(q, \epsilon)$ ($\epsilon$ the photon polarization vector) can be expressed in terms of a single hadronic parameter  $T_1^{B_{(s)} \to V}(0)$ for each  channel:
 \bea
 \Gamma(B_{(s)} \to V \, \gamma)&=&{C^2 \over 4 \pi} \left[T_1^{B_{(s)} \to V}(0)\right]^2 (m_b^2+m_s^2)\,M_{B_{(s)}}^3 \left( 1-{M_V^2 \over M_{B_{(s)}}^2} \right)^3 \,\,,\label{rate1} \\
 \Gamma(B \to K_2^* \, \gamma)&=&{C^2 \over 32 \pi} \left[T_1^{B \to VK_2^*}(0)\right]^2 (m_b^2+m_s^2)\,{M_B^5 \over M_{K_2^*}^2} \left( 1-{M_V^2 \over M_{B_{(s)}}^2} \right)^5 \,\,\, , \label{rate2}
 \eea
 where $C=4 \displaystyle{G_F \over \sqrt{2} } V_{tb} V_{ts}^* C_7^{eff} \displaystyle{e \over 16 \pi^2}$ and  the first equation applies both  to $B \to K^* \gamma$   and   $B_s \to \phi \gamma$.
 In the numerical analysis we set the particle masses and   lifetimes,  as well as  the CKM matrix elements to the PDG values;   for the quark masses we use $m_b \simeq 4.8$ GeV and $m_s \simeq 0.130$ GeV \cite{Colangelo:2000dp}. For the form factors we use results obtained by QCD sum rules \cite{Colangelo:2000dp}: in particular, for $B \to K^*$  we use the results obtained by three-point QCD sum rules \cite{Colangelo:1995jv}, based on the short-distance expansion,
  which provide  $T_1^{B \to K^*}(0)=0.38 \pm 0.06$, 
  and light-cone QCD sum rules (LCSR) \cite{Ball:2004rg}, based on  the light-cone expansion,   which give $T_1^{B \to K^*}(0)=0.333 \pm 0.028$. These values are larger than the lattice QCD result obtained in  quenched approximation \cite{Becirevic:2006nm}.
 LCSR calculations of $B \to K_2^*(1430)$ and $B_s \to \phi$ form factors give: $T_1^{B \to K^*_2}(0)=0.17 \pm 0.03 \pm 0.04$ \cite{Wang:2010ni} and
 $T_1^{B_s \to \phi}(0)=0.349 \pm 0.033 \pm 0.04$ \cite{Ball:2004rg}.

 In the case of  $\Lambda_b \to \Lambda \gamma$  we define the matrix elements
\be
<\Lambda(p^\prime,s^\prime)|{\bar
s}i \sigma_{\mu \nu} q^\nu b |\Lambda_b(p,s)> = {\bar u}_\Lambda \big[ f_1^T(q^2) \gamma_\mu +i
f_2^T(q^2)\sigma_{\mu \nu} q^\nu+f_3^T(q^2) q_\mu \big]
u_{\Lambda_b}  \label{lambda1}
\ee
\be
<\Lambda(p^\prime,s^\prime)|{\bar s}i \sigma_{\mu \nu} q^\nu \gamma_5 b
|\Lambda_b(p,s)> = {\bar u}_\Lambda
\big[ g_1^T(q^2) \gamma_\mu \gamma_5  +i g_2^T(q^2)\sigma_{\mu
\nu} q^\nu \gamma_5 +g_3^T(q^2) q_\mu \gamma_5 \big] u_{\Lambda_b}
  \label{lambda2}
\ee
with $u_\Lambda$ and $u_{\Lambda_b}$  the $\Lambda$  and $\Lambda_b$ spinors; $s$ denotes the baryon spin.
The determinations  of  the form factors in (\ref{lambda1})-(\ref{lambda2}) are quite uncertain. However, it is possible to  invoke
heavy quark symmetries for the hadronic matrix elements between an initial   spin=${1 \over 2}$ heavy baryon comprising a single heavy quark Q and a final   spin=${1 \over 2}$ light baryon;  due to the heavy quark symmetries    the
number of independent form factors is  two, since  for  $m_Q \to \infty$ and   a generic Dirac matrix $\Gamma$ one can write \cite{Mannel:1990vg}
\be <\Lambda(p^\prime,s^\prime)|{\bar s} \Gamma b
|\Lambda_b(p,s)> = {\bar
u}_\Lambda(p^\prime,s^\prime) \big\{ F_1(p^\prime \cdot v) + \spur{v} F_2  (p^\prime \cdot v)
\big\} \Gamma u_{\Lambda_b}(v,s)   \label{hqrelations} \ee
 where
$v={p \over M_{\Lambda_b}}$  is the $\Lambda_b$ four-velocity.
The form factors $F_{1,2}$ depend on the invariant  $\dd{p^\prime \cdot v= {M^2_{\Lambda_b}+M^2_\Lambda-q^2 \over 2 M_{\Lambda_b}}}$, but we refer to their $q^2$ dependence for convenience.
%
\begin{table*}[t]
\centering  \caption{PDG averages for the branching fractions  of radiative  $B$, $B_s$ and $\Lambda_b$  decay modes  \cite{pdg}.}\label{table:br}
\begin{tabular}{|c |c|}
\hline
mode &  {BR}   \\ \hline
$B^+ \to K^{*+} \, \gamma $ & \,\,\, $(42.1 \pm 1.8) \times 10^{-6}$\,\,\, \\ \hline
$B^0 \to K^{*0} \, \gamma $ & \,\,\,  $(43.3 \pm 1.5) \times 10^{-6}$ \,\,\, \\ \hline
\,\,\, $B^+ \to K^*_2(1430)^+ \, \gamma $\,\,\,  & \,\,\,  $(14 \pm 4) \times 10^{-6}$ \,\,\, \\ \hline
\,\,\, $B^0 \to K^*_2(1430)^0 \, \gamma $\,\,\,  & \,\,\,  $(12.4 \pm 2.4)\times 10^{-6}$\,\,\,  \\ \hline
${B}_s \to \phi \, \gamma $ & \,\,\,  $(57 \pm^{22}_{19})\times 10^{-6}$\,\,\,  \\ \hline
$\Lambda_b \to \Lambda \, \gamma $ & \hspace*{30pt} \,\,\,  $< 1.3 \times 10^{-3}$ \,\,\,  ($90\%$ C.L.) \\ \hline
\end{tabular}
\end{table*}
Using Eq.(\ref{hqrelations})  we can relate  the form factors in Eqs.(\ref{lambda1})-(\ref{lambda2}) to the functions $F_{1,2}$in Eq.(\ref{hqrelations}),
\bea  f_2^T&=&g_2^T=F_1 +{M_\Lambda \over M_{\Lambda_b}} F_2 \nonumber \\
f_1^T &=& g_1^T=q^2 {F_2 \over M_{\Lambda_b}}   \nonumber\\
f_3^T &=& -\Bigg(1-\frac{M_\Lambda}{M_{\Lambda_b}}\Bigg) \, F_2  \label{ffrelations}\\
g_3^T &=& \Bigg(1+ \frac{M_\Lambda}{M_{\Lambda_b}}\Bigg) \, F_2,
\nonumber \eea
at momentum transfer close to the maximum value $q^2 \simeq q^2_{max}=(M_{\Lambda_b}-M_\Lambda)^2$. 
We assume their validity  to  the whole phase space,  introducing a model dependence in the predictions.
The $\Lambda_b \to \Lambda \gamma$ decay width reads in terms of the form factors $F_1$ and  $F_2$:
\be
\Gamma(\Lambda_b \to \Lambda \, \gamma)={C^2 \over 4 \pi}   \left( F_1(0)+F_2(0)\,{M_{\Lambda} \over M_{\Lambda_b}} \right) \,(m_b^2+m_s^2)\,M_{\Lambda_b}^3 \left( 1-{M_{\Lambda}^2 \over M_{\Lambda_b}^2} \right)^3 \,\,.\label{rate3} \ee

A determination of  $F_1$ and  $F_2$ has been  obtained by three-point QCD sum rules in the $m_Q \to \infty$ limit  \cite{Huang:1998ek}. In  the following  we  use $F_1$, $F_2$ worked out in \cite{Colangelo:2007jy} updating some of the parameters
used  in \cite{Huang:1998ek}: $F_1(0) = 0.322 \pm 0.015$,  and $F_2(0) =-0.054 \pm  0.020$ \cite{Colangelo:2007jy}. Other determinations of the form factors in the transition $\Lambda_b \to \Lambda \gamma$ have been performed in \cite{Mannel:2011xg} at large recoil and in \cite{Gutsche:2013pp} in the covariant constituent quark model.
We set the $\Lambda_b$ and $\Lambda$ masses   and the $\Lambda_b$ lifetime to their PDG values \cite{pdg}.

The computed branching fractions are functions of $\displaystyle{1 \over R}$, through the dependence of the coefficient $C_7$ in the 6UED model.
The results are depicted in figs.\ref{plots1}, \ref{plots2},  where we have included the uncertainty on the form factor value at $q^2=0$, and a second uncertainty, intrinsic of the model,   due to the choice of the matching scale in the calculation of $C_7$ and of the  value of $N_{KK}$. These last uncertainty is discussed  in \cite{Freitas:2008vh}, together with another one which comes from fixing two boundary couplings $h_{1,2}$ which are ${\cal O}(1)$ and enter in the expression of the masses of the Higgs fields in this model. Altogether these uncertainties do not exceed $^{+17\%}_{-8\%}$ \cite{Freitas:2008vh},  and we include this range in our error on the branching fractions.
The experimental data for the various branching ratios are collected in Table \ref{table:br} and represent   PDG averages \cite{pdg}. For $B \to K^* \gamma$ modes, the results represent averages of   BaBar \cite{:2009we}, Belle \cite{Nakao:2004th} and CLEO  \cite{Coan:1999kh} measurements.  For $B \to K_2^*$,  the result is determined on the basis of the analysis in \cite{Aubert:2003zs}, while for $B_s \to \phi$ it stems from ref. \cite{:2007ni}. 
\begin{figure}[t!]
 \begin{center}
  \includegraphics[width=0.38\textwidth] {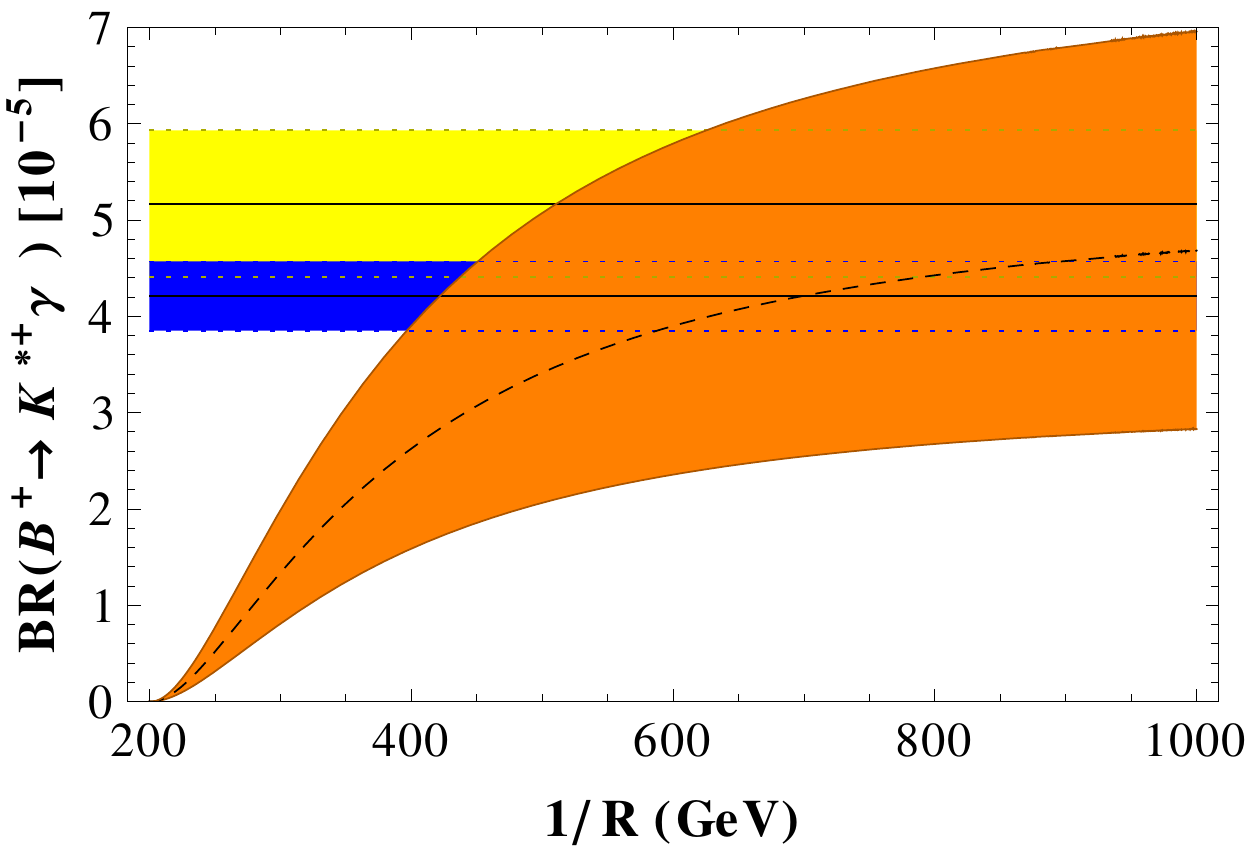} \hspace*{8pt}
  \includegraphics[width=0.38\textwidth] {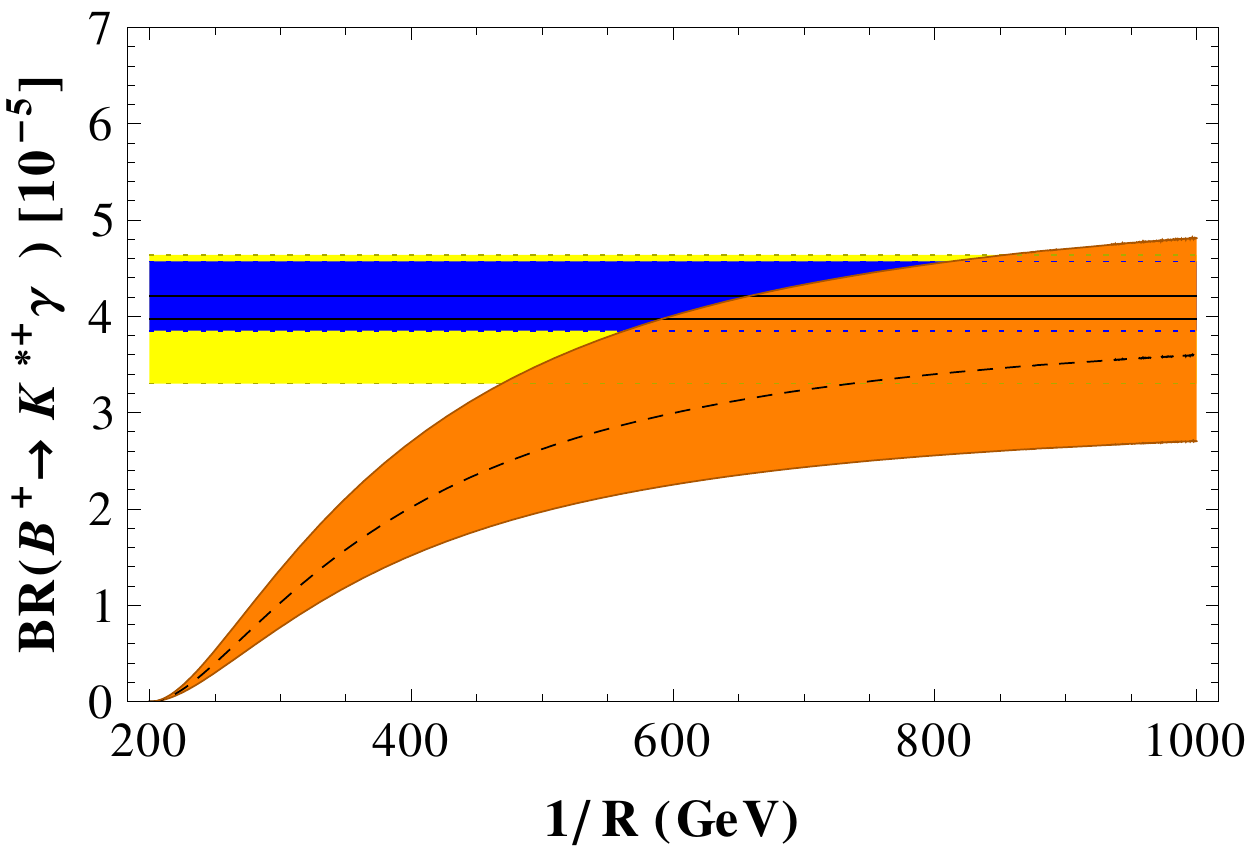}\\
   \includegraphics[width=0.38\textwidth] {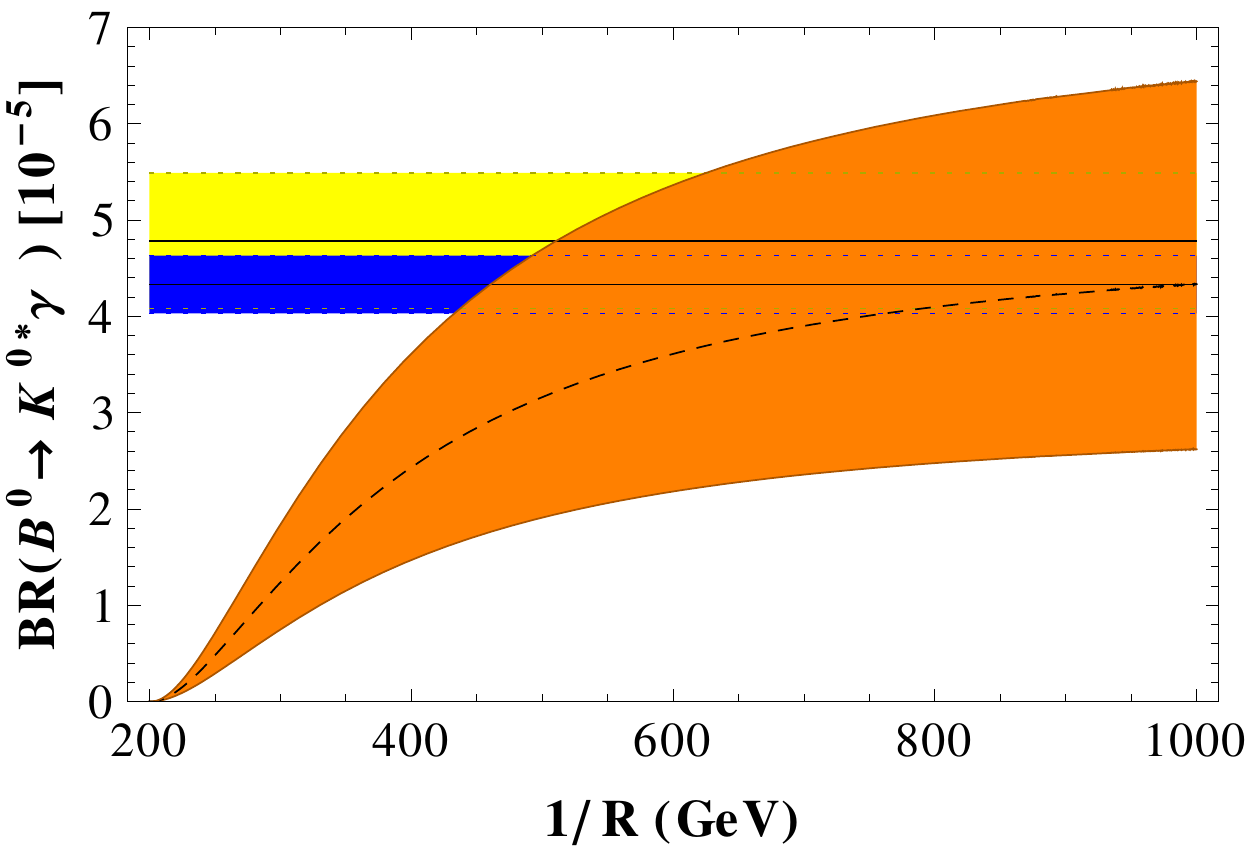} \hspace*{8pt}
    \includegraphics[width=0.38\textwidth] {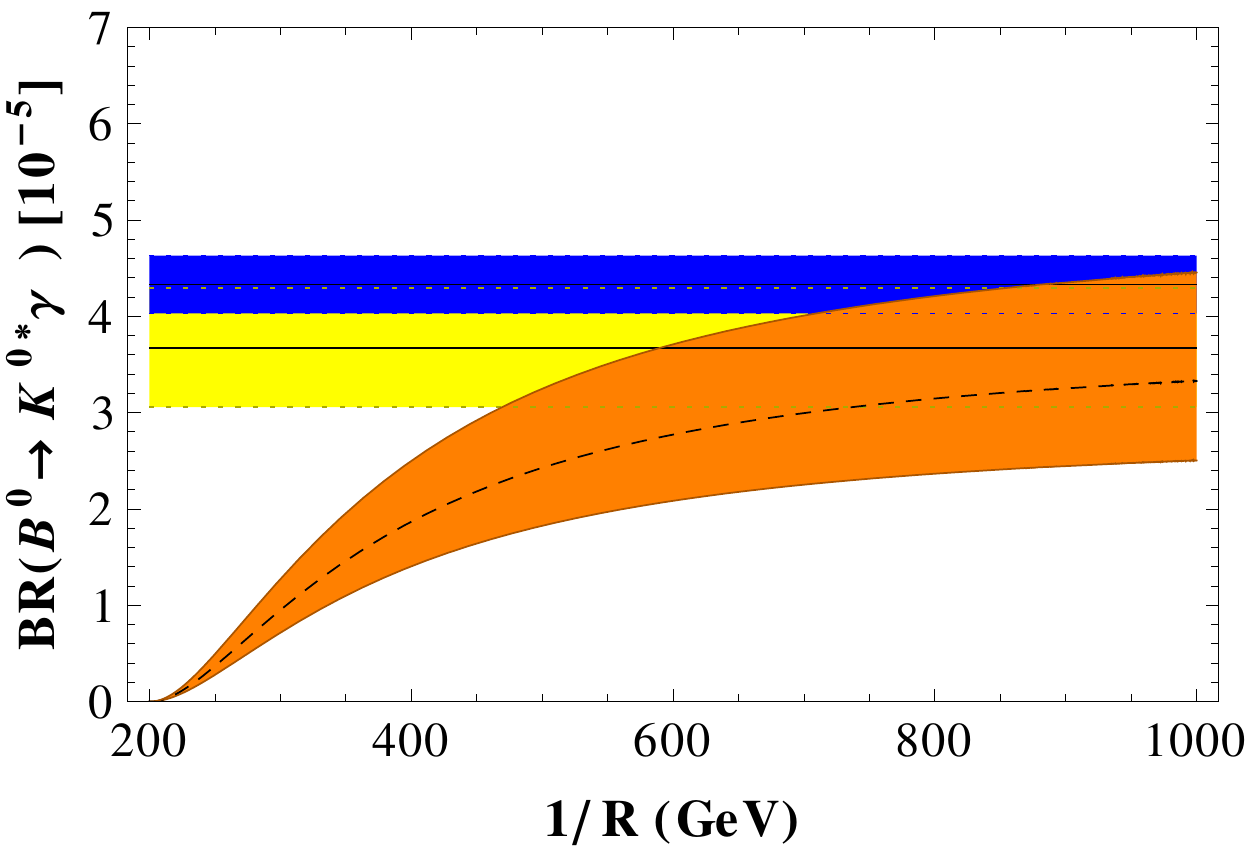}
\vspace*{0mm}
 \caption{Predicted branching fractions of  $B^+ \to K^{*+} \gamma$ (upper panels) and $B^0 \to K^{*0} \gamma$   (lower panels)  as a function of the compactification parameter $\displaystyle {1/R}$
(in units of $ \rm GeV$),  using the form factor in \cite{Colangelo:1995jv} (left) and in \cite{Ball:2004rg} (right).  The horizontal bands correspond to SM theoretical expectations with  $1$ $ \sigma$ uncertainties (yellow [light]) and experimental measurements with $2$ $ \sigma$ uncertainties (blue [dark]) respectively.}
  \label{plots1}
 \end{center}
\end{figure}
%
The upper bound on the ${\cal B}(\Lambda_b \to \Lambda \gamma)$ has been obtained in \cite{Acosta:2002fh}. Experimental data are represented as horizontal blue [dark] bands (at 95$\%$ c.l.) in fig.\ref{plots1},   and   in fig.\ref{plots2} for
 $B^0 \to K_2^{*0} \gamma$ (due the large experimental uncertainty, we only show  the neutral mode where the error  is  smaller) and  $B_s \to \phi \gamma$.

In the two upper plots in fig.\ref{plots1}  ${\cal B}(B^+ \to K^{*+} \gamma)$ is computed using either the form factor $T_1$ in \cite{Colangelo:1995jv} (left panel) or that derived in \cite{Ball:2004rg} (right panel). The same applies to the two lower plots, where the neutral channel $B^0 \to K^{*0} \gamma$ is considered.
There is a model dependence, with the resulting bounds :
$\displaystyle{1 \over R}\ge 397$ GeV (charged channel, form factors in \cite{Colangelo:1995jv}),
$\displaystyle{1 \over R}\ge 564$ GeV (charged channel, form factors in \cite{Ball:2004rg}),
$\displaystyle{1 \over R}\ge 433$ GeV (neutral channel, form factors in \cite{Colangelo:1995jv}),
$\displaystyle{1 \over R}\ge 710$ GeV (neutral channel, form factors in \cite{Ball:2004rg}).
As for  $B^0 \to K_2^{*0} \gamma$, we obtain  $\displaystyle{1 \over R}\ge 324$ GeV.
The other two plots in fig.\ref{plots2} refer to $B_s$ and $\Lambda_b$ decays. In a previous analysis \cite{Biancofiore:2012ij} also the decays $B \to K \eta^{(\prime)}\gamma$ has been considered in order to constrain UEDs models (with 1 and 2 extra dimensions respectively), finding for 6UED a lower bound of $\displaystyle{1 \over R} \ge 400$ GeV. 
\begin{figure}[h]
 \begin{center}
  \includegraphics[width=0.38\textwidth] {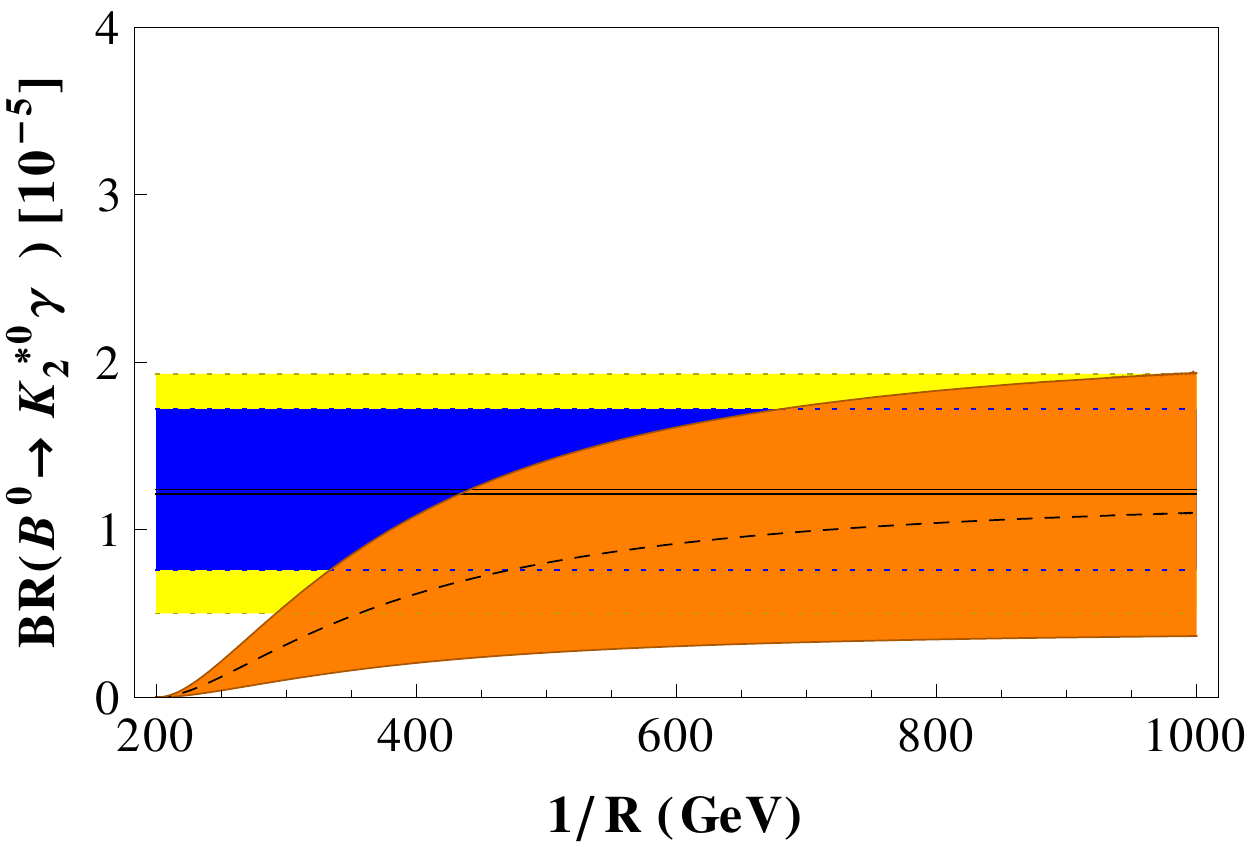}\\
  \includegraphics[width=0.38\textwidth] {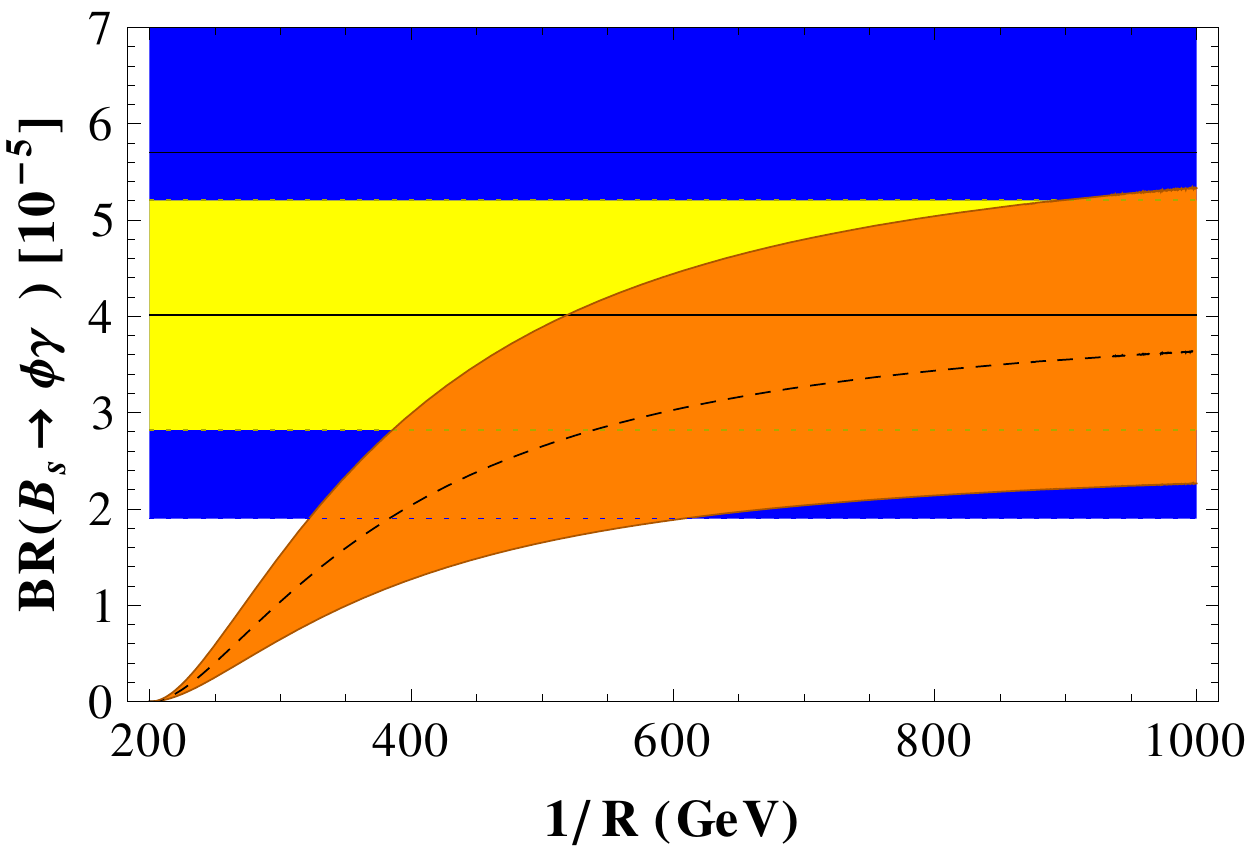}\hspace*{8pt}
   \includegraphics[width=0.38\textwidth] {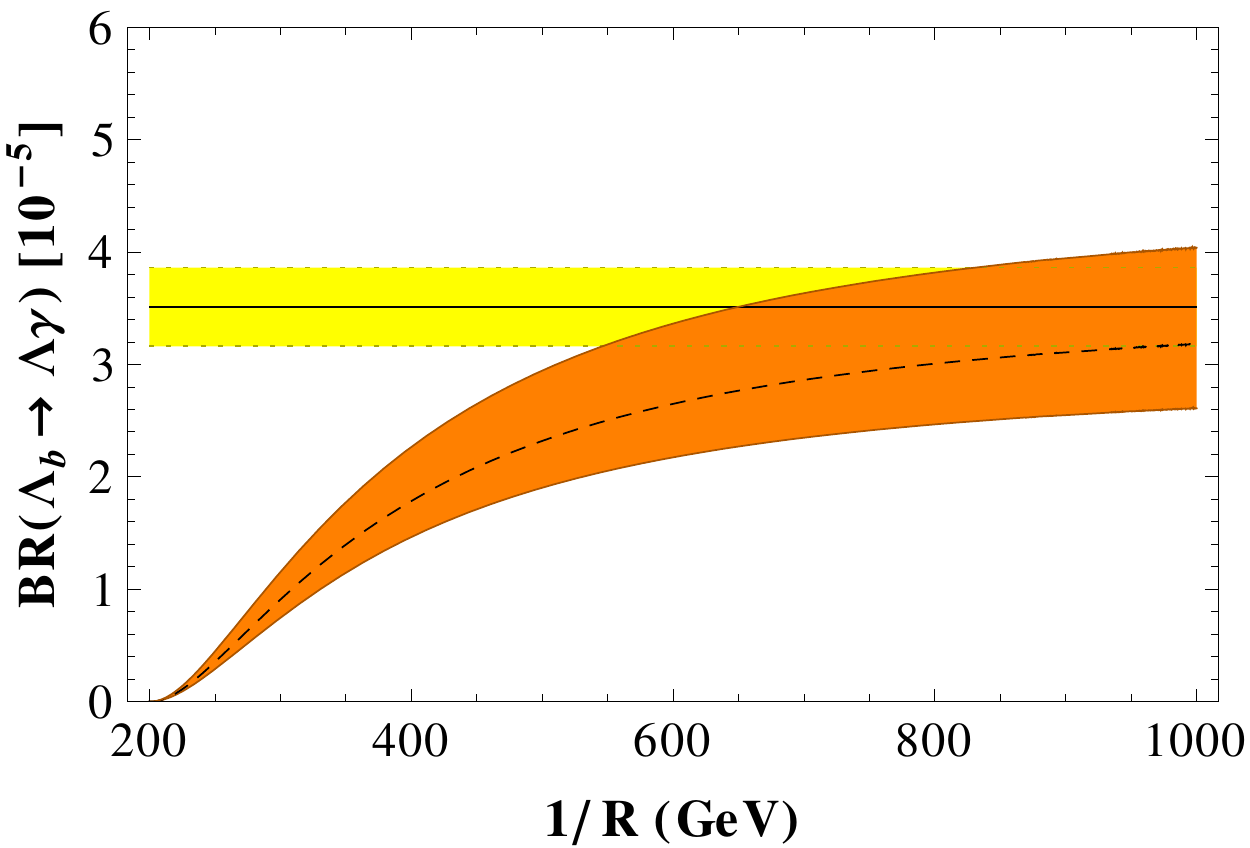}
\vspace*{0mm}
 \caption{Predicted branching fractions of  $B \to K^*_2 \gamma$,  $B_s \to \phi \gamma$ and $\Lambda_b \to \Lambda \gamma$ as a function of the compactification parameter $\displaystyle{1/R}$ (in units of $\rm GeV$).
The SM results (with $1$ $\sigma$ uncertainties) and experimental measurements (with $2 \sigma$ uncertainties) are represented as horizontal yellow [light] and blue [dark] bands, respectively. }
  \label{plots2}
 \end{center}
\end{figure}

Bounds from direct searches of KK modes have been discussed in \cite{Dobrescu:2007xf}, where the hadron collider phenomenology of (1,0) KK modes in the 6UED model was studied. The  limit $\displaystyle{1 \over R}\ge 270$ GeV  was found, set  from  direct searches at the  Tevatron.
Our  bound is  more restrictive than the one obtained from the inclusive radiative $B$ transition \cite{Freitas:2008vh}, as in the case of  a single UED. 
Since the discovery signals of  UEDs at the CERN LHC or at a future $e^+ e^-$ linear collider,  searching for leptons plus missing energy,  are expected with  the experimental reach  of about
$\displaystyle{1/R} \simeq 1$ TeV \cite{Ghosh:2008ji,Bhattacherjee:2010vm,Choudhury:2011jk}, the complementarity of  the studies of indirect effects of possible UEDs in flavour observables is noticeable.

\section*{Acknowledgements}
I thank P. Colangelo and F. De Fazio for the valuable collaboration. This work is supported in part by the Italian Miur Prin 2009.

\end{document}